\documentclass[12pt,preprint]{aastex}
\usepackage{textcomp}
\usepackage{rotating}
\usepackage{hyperref}
\usepackage[all]{hypcap}
\hypersetup{
    bookmarks=true,         
    unicode=false,          
    pdftoolbar=true,        
    pdfmenubar=true,        
    pdffitwindow=false,     
    pdfstartview={FitH},    
    pdftitle={My title},    
    pdfauthor={Author},     
    pdfsubject={Subject},   
    pdfcreator={Creator},   
    pdfproducer={Producer}, 
    pdfkeywords={keyword1} {key2} {key3}, 
    pdfnewwindow=true,      
    colorlinks=true,        
    linkcolor=blue,         
    citecolor=blue,         
    filecolor=magenta,      
    urlcolor=cyan           
}

\slugcomment{}

\shorttitle{lid removal mechanism}
\shortauthors{Joshi et al.}

\begin{document}


\title{Sequential Lid Removal in a Triple-Decker Chain of CME-Producing Solar Eruptions}

\author{Navin Chandra Joshi\altaffilmark{1,2}, Alphonse C. Sterling\altaffilmark{3}, Ronald L. Moore\altaffilmark{3,4}, Bhuwan Joshi\altaffilmark{1}}

\altaffiltext{1}{Udaipur Solar Observatory, Physical Research Laboratory, Udaipur 313 001; navinjoshi@prl.res.in, njoshi98@gmail.com}
\altaffiltext{2}{Department of Physics, SRM University, Delhi-NCR, Sonepat-131029, Haryana, India; ncjoshi@srmuniversity.ac.in}
\altaffiltext{3}{NASA Marshall Space Flight Center, Huntsville, AL 35812, USA}
\altaffiltext{4}{Center for Space Plasma and Aeronomic Research (CSPAR), UAH, Huntsville, AL 35805, USA}


\begin{abstract}

We investigate the onsets of three consecutive coronal mass ejection (CME) eruptions in 12 hours from a large bipolar active region (AR) observed by \textit{SDO}, \textit{STEREO}, \textit{RHESSI}, and \textit{GOES}. Evidently, the AR initially had a ``triple--decker" configuration: three flux ropes in a vertical stack above the polarity inversion line (PIL).  Upon being bumped by a confined eruption of the middle flux rope, the top flux rope erupts to make the first CME and its accompanying AR--spanning flare arcade rooted in a far-apart pair of flare ribbons.  The second CME is made by eruption of the previously-arrested middle flux rope, which blows open the flare arcade of the first CME and produces a flare arcade rooted in a pair of flare ribbons closer to the PIL than those of the first CME.  The third CME is made by blowout eruption of the bottom flux rope, which blows open the second flare arcade and makes its own flare arcade and pair of flare ribbons.  Flux cancellation observed at the PIL likely triggers the initial confined eruption of the middle flux rope.  That confined eruption evidently triggers the first CME eruption.  The lid-removal mechanism instigated by the first CME eruption plausibly triggers the second CME eruption.   Further lid removal by the second CME eruption plausibly triggers the final CME eruption.

\end{abstract}




\section{Introduction}
\label{sec1}

Eruptions of solar filaments/prominences is among the most interesting phenomena that occurs on the Sun. Solar filaments are cool and relatively dense plasma material suspended in the hot and relatively tenuous solar corona. Prominences and filaments are the same objects, being called filaments when observed against the solar disk and prominences when viewed at the solar limb \citep[see reviews by][and references cited therein]{Labrosse10,Mackay10,Parenti14,Gibson18}. According to flux-rope models of filaments and prominences, they are supported by the helical magnetic field of the flux rope \citep[e.g.,][and references cited therein]{Filippov15}.

Filaments/prominences and flux ropes remain stable in the solar corona by the equilibrium among the upward magnetic pressure, downward magnetic tension, and gravity. They erupt when this equilibrium is disrupted. Various theories/models have been proposed in order to explain the onset of solar eruptions \citep{Forbes00,Lin03,Schmieder13}. Three well-known models that have been proposed are ``tether--cutting," ``flux cancellation," and the ``magnetic breakout" models. The tether-cutting scenario was suggested by \cite{Moore92} and \cite{Moore01}. According to this model, low-lying tether-cutting magnetic reconnection triggers and unleashes the eruption. The low-lying sheared arcade reconnects successively and forms the flux rope. Continuing low-lying reconnection reduces the tension and allows the eruption of the formed flux rope. The flux rope then erupts, and reconnection occurs beneath it. Several observed events have been studied and interpreted using this model \citep[e.g.,][]{Liu12a,Liu13a,Joshi14,Joshi2017}. The magnetic breakout model considers a quadrupolar magnetic configuration  \citep{Antiochos99}. According to this model, reconnection at the magnetic null point that lies on top of the middle arcade triggers the eruption of the underlying flux rope. Fast reconnection occurs below the erupting flux rope in the next stage \citep[see][]{Karpen12}. Various signatures of the breakout model have been observed and interpreted in solar eruptive events \citep{Aulanier00,Gary04,Sterling04b,Sterling04a,Deng05,Joshi07,Lin10,Aurass11,Aurass13,Mitra18}. According to the flux-emergence model suggested by \cite{Chen00}, an emerging bipole within the filament channel cancels with the magnetic field below the filament (or the flux rope) and triggers its eruption. Apart from this, ideal MHD instability has also been investigated to trigger solar eruptions, e.g., by catastrophic loss of equilibrium \citep[e.g.,][]{Forbes91}, kink instability \citep[e.g.,][]{Torok04,Torok05}, and torus instability \citep[e.g.,][]{Kliem06}. 

Triggering the eruption of a filament/flux rope by a nearby erupting filament and/or flux rope has also been observed and discussed. Such eruptions and called sympathetic eruptions \citep{Wang01,Wang07,Liu09,Schrijver11,Jiang11,Yang12,Shen12,Joshi16b,Wang16}. \cite{Wang01} studied sympathetic flares and accompanying CMEs from two active regions (ARs), NOAA ARs 8869 and 8872, and discussed 
the linkage between them through connecting loops. \cite{Liu09} analyzed successive solar eruptions and CMEs that occurred on 2005 September 13 and found an interrelation among the eruptions. A series of eruptions, CMEs, and related events were observed over 2010 August 1-2 and it was reported that they were connected by a system of separatrices, separators and quasi-separatrix layers \citep{Schrijver11}. Successive eruptions that occurred on 2003 November 19 have been interpreted as being sympathetic in nature using coronal dimming observations \citep{Jiang11}. \cite{Yang12} investigated successive sympathetic eruptions of two filaments in a bipolar helmet streamer magnetic configuration observed on 2005 August 5. In a typical quadrupolar magnetic configuration, \cite{Shen12} observed sympathetic partial and full eruptions that occurred on 2011 May 12. Recently, \cite{Joshi16b} studied two successive adjacent filament eruptions and investigated the chain of reconnections during them. \cite{Wang16} investigated the sympathetic nature of two eruptions that occurred on 2015 March 15 and also noted that the second eruption gave the largest geomagnetic storm of the current solar cycle. Two successive flux rope eruptions and accompanying CMEs that occurred on 2012 January 23 have been studied by several authors \citep{Joshi13a,Cheng13,Sterling14}.  Some sort of magnetic/physical connections between the sympathetic eruptions have been observed in all these studies. For example, \cite{Sterling14} argued that removal of flux above a filament/flux-rope-carrying non-potential field arcade can lead to destabilization and eruption of the filament/flux rope, a process they called {\it lid removal}. A few attempts to model these kinds of sympathetic eruptions have also been made \citep[see e.g.,][]{Torok11,Lynch13}.

The eruption of double-decker filaments, which are composed of two filament branches separated in height, have 
been observed and studied \citep[][]{Su11,Liu12b,Kliem14,Ding14,Zhu14,Cheng14a,Awasthi19,Zheng19}.
\citet[][]{Liu12b} observed the eruption of a double-decker filament having its pre-eruption two branches separated in height 
by about 13 Mm, and found evidence that transfer of flux and current from the lower branch to the upper branch was 
the triggering process for the partial eruption of the upper branch \citep[see also][]{Kliem14}.
\citet[][]{Zhu14} studied a similar kind of partial eruption of the upper branch of a double-decker filament that occurred on 2012 May 9. They interpret the eruption in the same way as suggested by \citet[][]{Liu12b}.

Several works have investigated an interesting set of multiple eruptions occurring from active region AR 11745.  \citet[][]{Li13a} studied the dynamics and homologous eruption of four flux ropes from AR 11745 on 2013 May 20-22, in order to understand the characteristics of flux ropes. They observed that these flux ropes erupted consecutively from the active region.  Slow flux rope eruption did not produce a CME, while fast eruption did produce a CME\@. \citet[][]{Cheng14b} studied the morphological evolution, kinematics and thermal properties of one of the erupting magnetic flux ropes on 2013 May 22 from AR 11745. For that particular flux rope eruption, they proposed two distinct phases of evolution. In the first phase, the slow-rising phase of the eruption, reconnection occurred within the quasi-separatrix layers surrounding the flux rope. In the later phase, the impulsive-acceleration phase of the eruption, fast reconnection occurred underneath the flux rope. The torus instability was found to be responsible for the transition from slow to fast erupting phase. \citet[][]{Makela16} studied the source region of type II radio bursts observed on 2013 May 22, and found that the interaction of two CMEs launched from the AR 11745 coincided with the type II emissions. They further recognized that successive eruptions from the same active region produced CME--CME interaction that resulted in solar energetic particle acceleration \citep[see also][]{Ding14}. None of these studies however focused on the triggering process for the successive flux rope eruptions from above the same polarity inversion line (PIL) of AR 11745. 

In this work, we revisit the eruption events that occurred in AR 11745 on 2013 May 22, with the motivation of exploring the complex triggering process. We present significant additional results for the triggering processes of these four successive flux rope eruptions from the single PIL of the bipolar active region NOAA AR 11745. The first eruption is of the ``confined" variety, during which a filament gets activated and attains quasi-stable height. The three subsequent eruptions are ejective eruptions (blowout eruptions) that each produce a CME. We interpret that the three consecutive successful eruptions are intimately linked with a ``triple--decker" flux rope configuration formed in the source region. We focus our study on the cause of onset and the interrelationship among these four eruptions, including evidence that sequential lid removal triggers the second and third CMEs. In Section~\ref{sec2}, we describe the observational data set. Section~\ref{sec3} presents the morphology of the active region and the potential--field source--surface (PFSS) extrapolation. The observed onsets and dynamics of the four eruptions and their CMEs are presented in Section~\ref{sec4}. Section~\ref{sec5} summarizes 
the main results and presents additional discussion.

\begin{figure}[!ht]
\vspace*{0cm}
\centerline{
	\hspace*{0\textwidth}
	\includegraphics[width=1\textwidth,clip=]{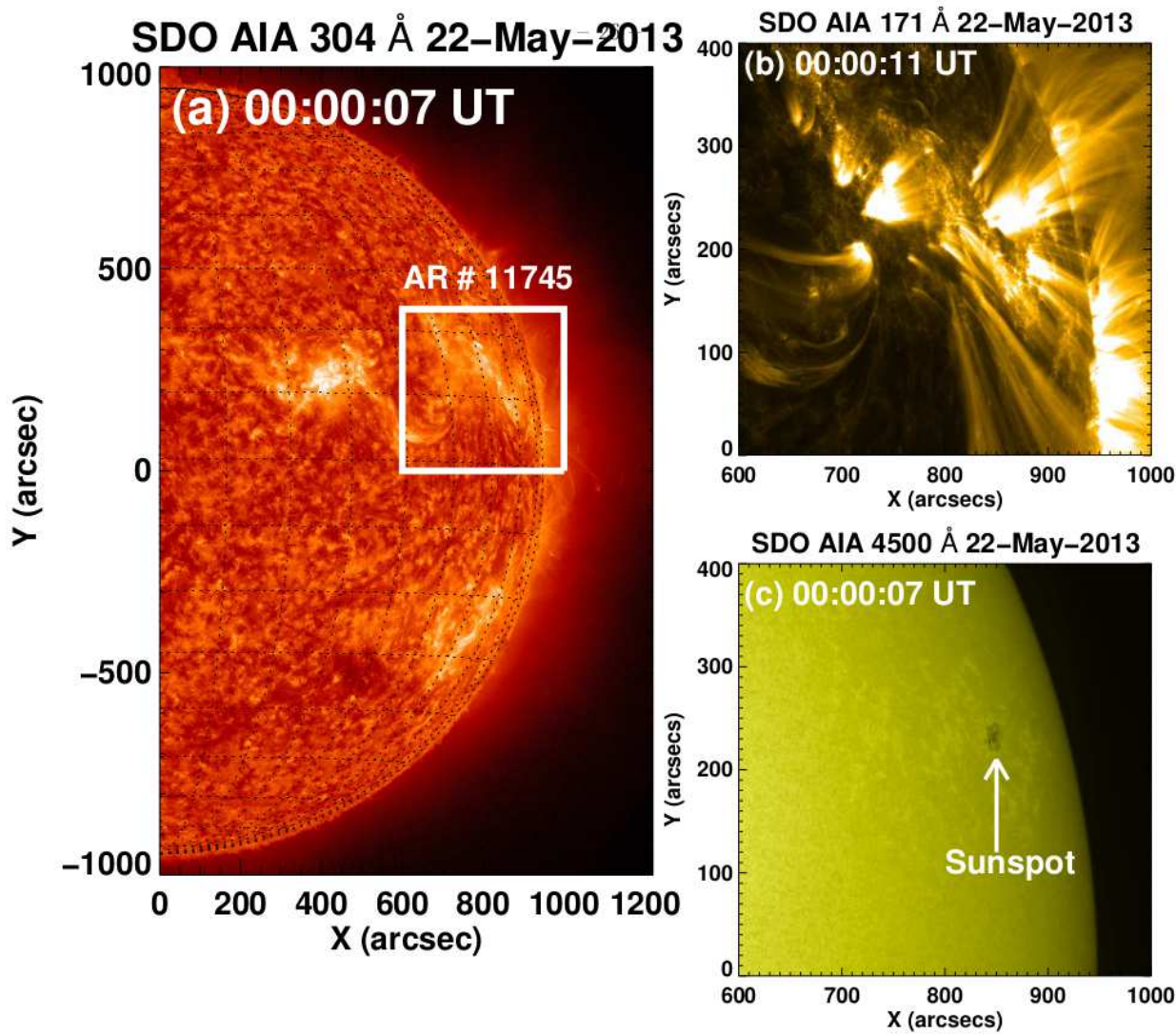}
	}
\vspace*{0cm}
\caption{(a) \textit{SDO}/AIA 304 \AA\ image at 00:00:07 UT on 2013 May 22, showing the western half of the Sun. The white box covers the active region area and is the field--of--view of panels (b) and (c). (b) \textit{SDO}/AIA 171 \AA\ image at 00:00:11 UT on 2013 May 22. (c) \textit{SDO}/AIA 4500 \AA\ continuum image at 00:00:07 UT on 2013 May 22. A sunspot of the region is shown by the white arrow in panel (c).}
\label{fig1}
\end{figure}


\section{Data set}
\label{sec2}

Data from various instruments are used to study the four consecutive eruptive events. Ultraviolet (UV) and Extreme Ultraviolet (EUV) images are observed by the Atmospheric Imaging Assembly \citep[AIA;][]{Lem12}, which is an instrument on board the {\it Solar Dynamics Observatory} \citep[\textit{SDO};][]{Pesnell12}. AIA observes the Sun with EUV (304, 131, 171, 193, 211, 335, and 94 \AA), UV (1600 and 1700 \AA), and white light (4500 \AA) channels with a temporal cadence of 12 s, 24 s and 1 hr, respectively and pixel size of $\rm 0.6\arcsec$.  Line-of-sight (LOS) magnetograms with temporal cadence of 45 s with pixel size of $\rm 0.5\arcsec$ are obtained by SDO's Helioseismic and Magnetic Imager \citep[HMI;][]{Schou12}. Full--Sun X--ray flux at 1-8 \AA\ and 0.5-4 \AA\ is observed by {\it the Geostationary Operational Environmental Satellite} (GOES)\@. Light curves and images in X-rays for the energy range 3-100 keV are obtained from the {\it Reuven Ramaty High Energy Solar Spectroscopic Imager} \citep[\textit{RHESSI};][]{Lin02}. For RHESSI X--ray image construction, we use the PIXON algorithm with a time integration time of 20 s.

We also use imaging data from the Sun Earth Connection Coronal and Heliospheric Investigation/Extreme Ultraviolet Imager (SECCHI/EUVI), which is an instrument on-board the {\it Solar Terrestrial Relations Observatory} \citep[\textit{STEREO}][]{Wuelser04,Howard08}. It observes the Sun in four wavelengths, viz.\ 304, 195, 284, and 171 \AA, and provides images with 1.6\arcsec\ pixels.
To study the CMEs that accompany the eruptions, we use coronagraph data from {\it STEREO-A}/COR1 \citep{Howard08} and the Large Angle and Spectrometric Coronagraph (LASCO; Brueckner et al. 1995) on board the {\it Solar and Heliospheric Observatory (SOHO)}.


\section{Morphology of the active region and PFSS extrapolation}
\label{sec3}

Figure~\ref{fig1}(a) shows the western half of the Sun in an \textit{SDO}/AIA 304 \AA\ image at $\approx$00:00 UT on 2013 May 22. The bipolar active region from which the eruptions originated is shown in the white box in Figure~\ref{fig1}(a). As seen, the active region was in the north-west quadrant of the Sun. Figures~\ref{fig1}(b) and~\ref{fig1}(c) are zoomed views of the active region in 
171 \AA\ and 4500 \AA\ images. A sunspot can be seen in the 4500 \AA\ image (shown by the white arrow in Figure~\ref{fig1}(c)).

\begin{figure}[!ht]
\vspace*{0cm}
\centerline{
	\hspace*{0.03\textwidth}
	\includegraphics[width=1\textwidth,clip=]{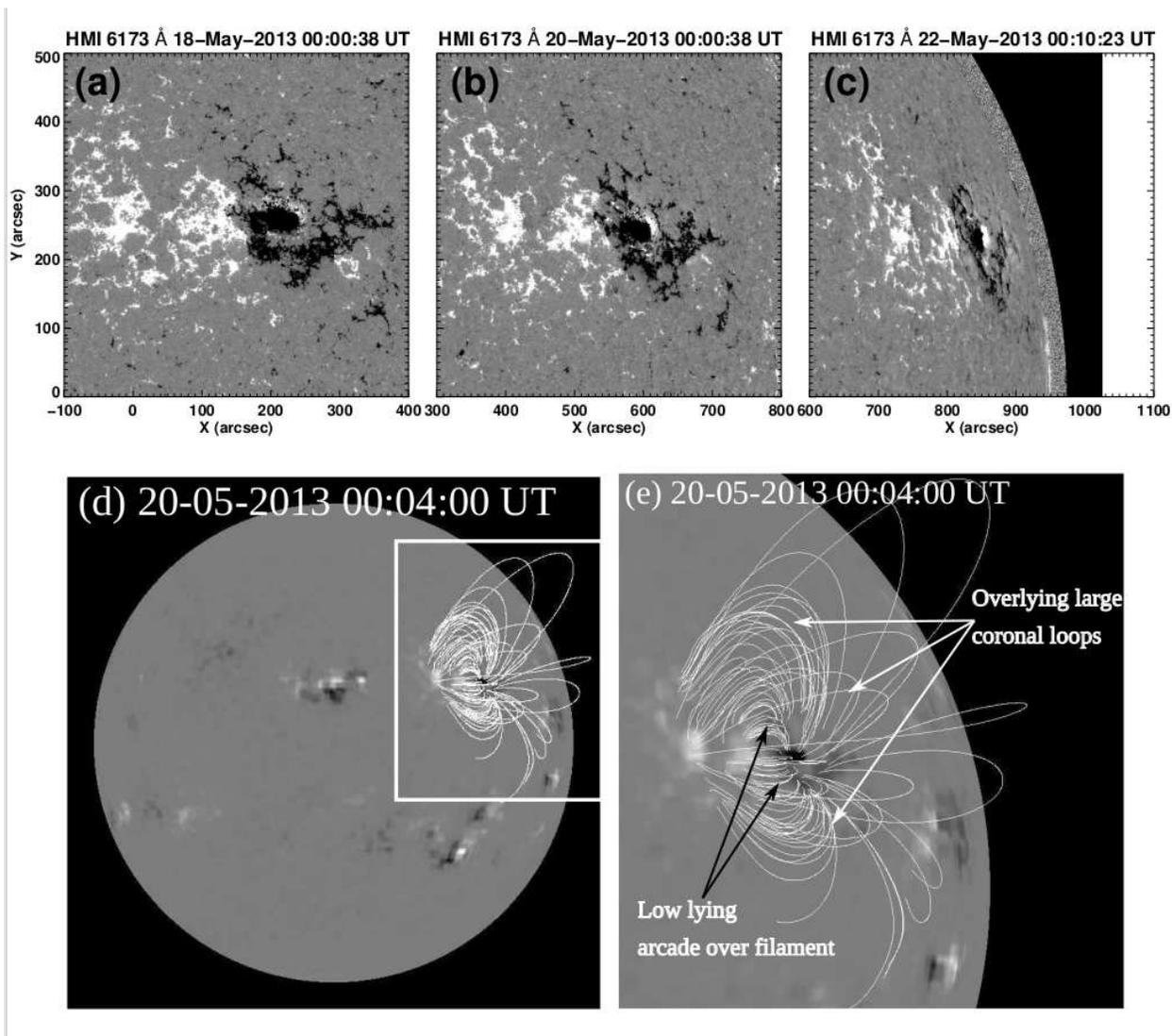}
    	}
\vspace*{0cm} 
\caption{Upper panel ((a)--(c)): \textit{SDO}/HMI line--of--sight photospheric magnetograms on 2013 May 18--22. Bottom panel (d): Potential--field source--surface (PFSS) extrapolated magnetic field lines on a full disk \textit{SDO}/HMI magnetogram, showing the coronal magnetic field structure over the active region. Bottom panel 
(e): Zoomed view corresponding to the white box drawn in the panel (d). An animation of the \textit{SDO}/HMI 720 s line--of--sight magnetograms (panel (c)) showing magnetic flux cancellation at the AR\textquotesingle s main PIL from $\approx$00:00 UT to  $\approx$14:00 UT on 2013 May 22 is available in the online journal.}
\label{fig2}
\end{figure}

Figures~\ref{fig2}(a)--(c) show \textit{SDO}/HMI line--of--sight magnetograms during the period 2013 May 18--22. The magnetic field structure of the active region NOAA AR 11745 is overall bipolar, with leading negative and following positive polarities. There was a large negative-polarity sunspot in this active region. On comparing the HMI images from 2013 May 18 to 22, we see that the overall magnetic structure of the region does not show much change. However, from the HMI 720 s line-of-sight magnetogram movie clearly indicates that cancellation is occurring at the main neutral line (see the movie accompanying Figure~\ref{fig2}).  We have also inspected the HMI SHARP series vector magnetograms, and they are consistent with cancellation occurring at that neutral line. We only display the line-of-sight movie here because the vector movies are of poorer fidelity. From the observed apparent flux cancellation we can now conclude that the build-up and triggering mechanism of the earliest confined flux-rope eruption was probably flux cancellation at the PIL \citep[][]{Moore92}. 

The field structure in the active region's corona is obtained using the potential-field source-surface \citep[PFSS,][]{Schrijver03} extrapolation method. PFSS is an IDL--based algorithm that is available in the SolarSoftWare package (SSW\footnote{\url{http://www.lmsal.com/solarsoft/}}). The extrapolated magnetic field lines are shown in Figures~\ref{fig2}(d)--(e). Figure~\ref{fig2}(d) shows a full disk magnetogram of 2013 May 20 at $\approx$00:04 UT\@. Figure~\ref{fig2}(e) shows a zoomed view corresponding to the  white box area shown in Figure~\ref{fig2}(d). From the extrapolation we can see a low-lying arcade, as well as overlying larger coronal-field loops, connecting the negative and positive polarities across the PIL.


\section{Dynamic evolution of the successive eruptions}
\label{sec4}


\subsection{Investigation of intensity profiles}
\label{sec4.1}

\begin{figure}[!ht]
\vspace*{-3cm}
\centerline{
	\hspace*{0.02\textwidth}
	\includegraphics[width=1\textwidth,clip=]{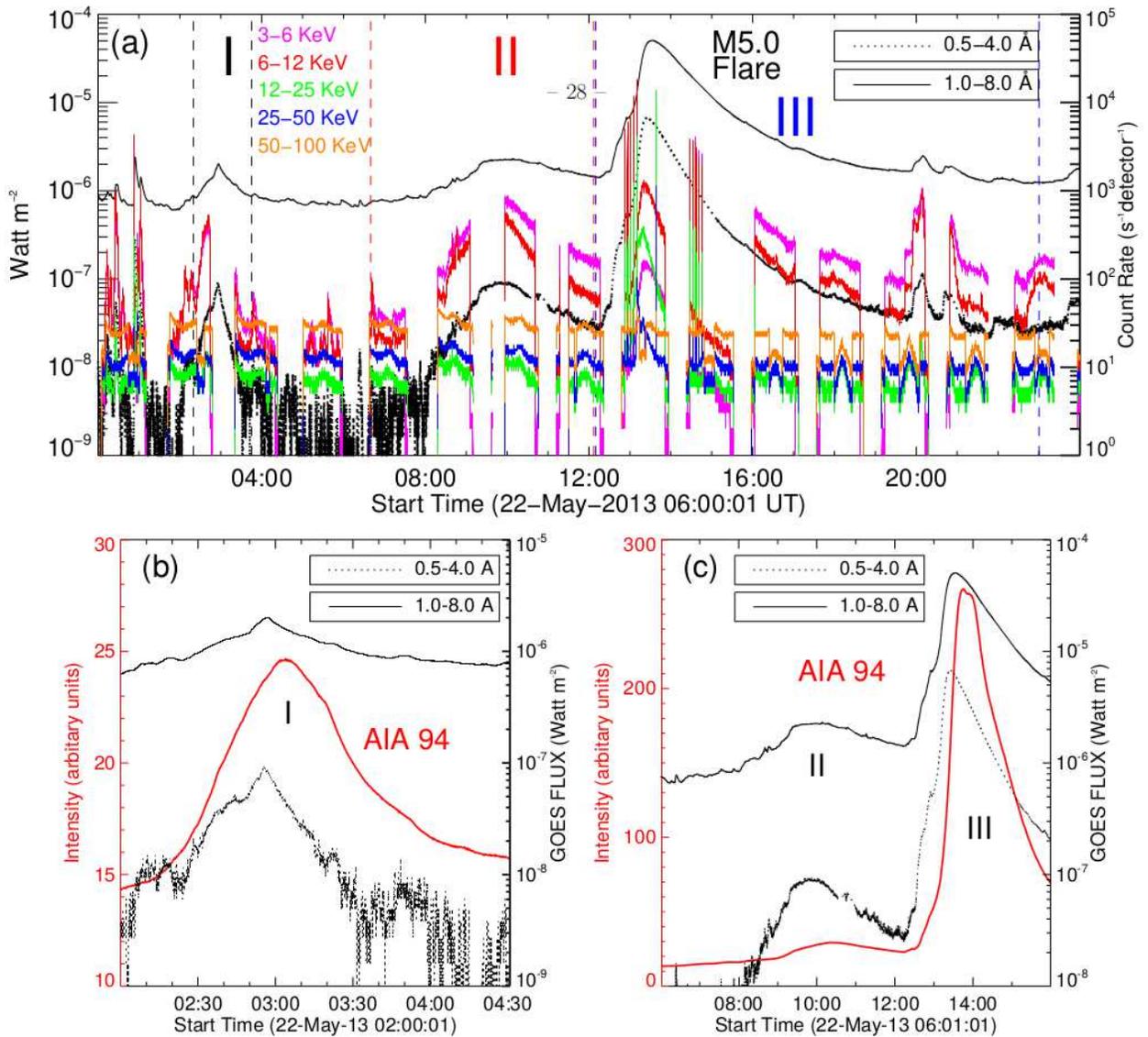}
	}
\vspace*{0cm}
\caption{Upper panel (a): {\it GOES} and {\it RHESSI} X--ray temporal flux profiles from $\approx$00:00 UT to $\approx$24:00 UT on 2013 May 22. Spikes in the RHESSI time profile, mainly during the M5.0 flare, are artifacts due to changes of the RHESSI attenuator state. Bottom panel (b): \textit{SDO}/AIA 94 \AA\ intensity profile (red) from $\approx$02:00 UT to $\approx$04:30 UT on 2013 May 22, during the confined--flare filament eruption that triggered the first CME eruption. Bottom panel (c): \textit{SDO}/AIA 94 \AA\ intensity profile (red) from $\approx$05:30 UT to $\approx$16:00 UT on 2013 May 22, during the second and third CME eruptions and their flares. The GOES X--ray fluxes at 1.0--8.0 \AA\ (solid black) and 0.5--4.0 \AA\ (dotted black) are also overplotted for comparison in panels (b) and (c).}
\label{fig3}
\end{figure}

The GOES and RHESSI X--ray temporal flux profiles are shown in Figure~\ref{fig3}(a). We can see three events of flux enhancements (marked by I, II, and III in Figure~\ref{fig3}(a)) that correspond to three CME--producing eruptive flares. The first event (i.e. event I) is actually a composite of two distinct eruptive events that include an early confined eruption of an activated filament that subsequently triggered the eruption of an overlying flux rope system, causing the first CME. The first event, which starts to rise at $\approx$02:25 UT, peaks at $\approx$02:56 UT, and ends at $\approx$03:08 UT, is a small C1.9-class flare. The second event, which starts at  $\approx$06:40 UT, peaks at $\approx$10:08 UT, and ends at $\approx$12:00 UT, is a C2.3-class flare. The third event (which is an M5.0 class flare) starts at $\approx$12:20 UT, peaks at $\approx$13:32 UT, and ends at $\approx$23:00 UT, displaying an extended decay phase. The RHESSI X-ray profiles also show peaks (see colored profiles in Figure~\ref{fig3}(a)) that correspond to the peaks in the GOES X-ray profiles during the flares. In the RHESSI X--ray profile, we can observe many large data gaps spanning tens of minutes. These gaps are due periods of RHESSI night or passage through the South Atlantic Anomaly (SAA).

To confirm the source of X--ray flux enhancements during these three eruption episodes, we also made AIA 94 \AA\ time profiles (shown by solid red line in Figures~\ref{fig3}(b)--(c)). Figure~\ref{fig3}(b) shows the intensity profile from $\approx$02:00 UT to $\approx$04:30 UT during the confined-eruption flare, while Figure~\ref{fig3}(c) shows the intensity profile from $\approx$06:00 UT to $\approx$16:00 UT during the flares of the second and third CME eruptions. The area used to calculate the intensity profiles during the first eruption is the field--of--view (FOV) of Figure~\ref{fig4}(c), while for the second and third eruptions the area used is the FOV of Figure~\ref{fig7}(a). The average values within these areas are used for these intensity-time profiles. Here we see that the variation in AIA 94 \AA\ flux is similar to that of GOES X-ray flux (cf.\ AIA and GOES flux profiles in Figures~\ref{fig3}(b)--(c)). This confirms that most of the X-ray flux is coming from AR~11745 during theses three flare events.


\subsection{Onset and progression of the first CME eruption}
\label{sec4.2}

\begin{figure}[!ht]
\vspace*{0cm}
\centerline{
	\hspace*{0.0\textwidth}
	\includegraphics[width=1\textwidth,clip=]{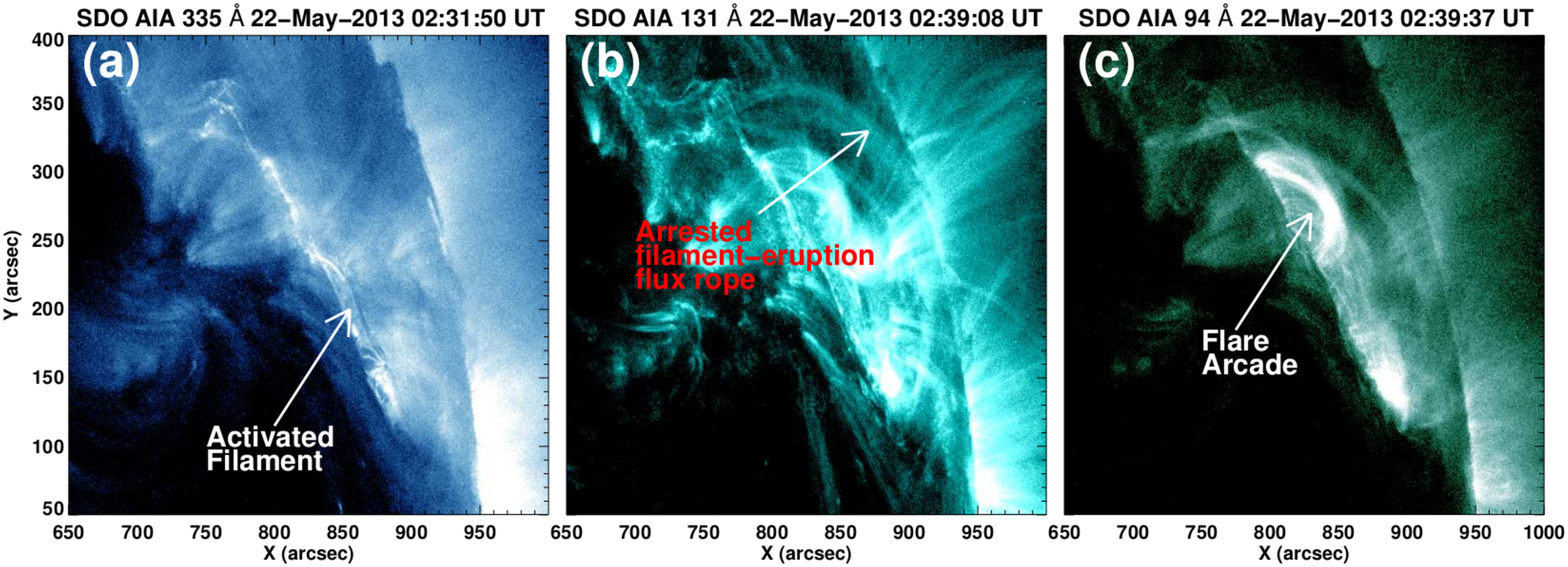}
	}
\vspace*{0cm}
\caption{Images near the time of the first eruption episode.  (a) \textit{SDO}/AIA 335 \AA\ image at $\approx$02:31 UT showing the activated filament. (b) \textit{SDO}/AIA 131 \AA\ image at $\approx$02:39 UT, showing the arrested filament-eruption flux rope. (c) \textit{SDO}/AIA 94 \AA\ image at $\approx$02:39 UT, showing the flare arcade made by the confined filament eruption. Animations of the \textit{SDO}/AIA 335 \AA\ intensity (animation 1) and running difference images of \textit{SDO}/AIA 94 (animation 2) showing the activation of filament and subsequent eruption from $\approx$02:15 UT to $\approx$03:15 UT on 2013 May 22 are available in the online journal.}
\label{fig4}
\end{figure}

\begin{figure}[!ht]
\vspace*{0cm}
\centerline{
	\hspace*{0.0\textwidth}
	\includegraphics[width=1\textwidth,clip=]{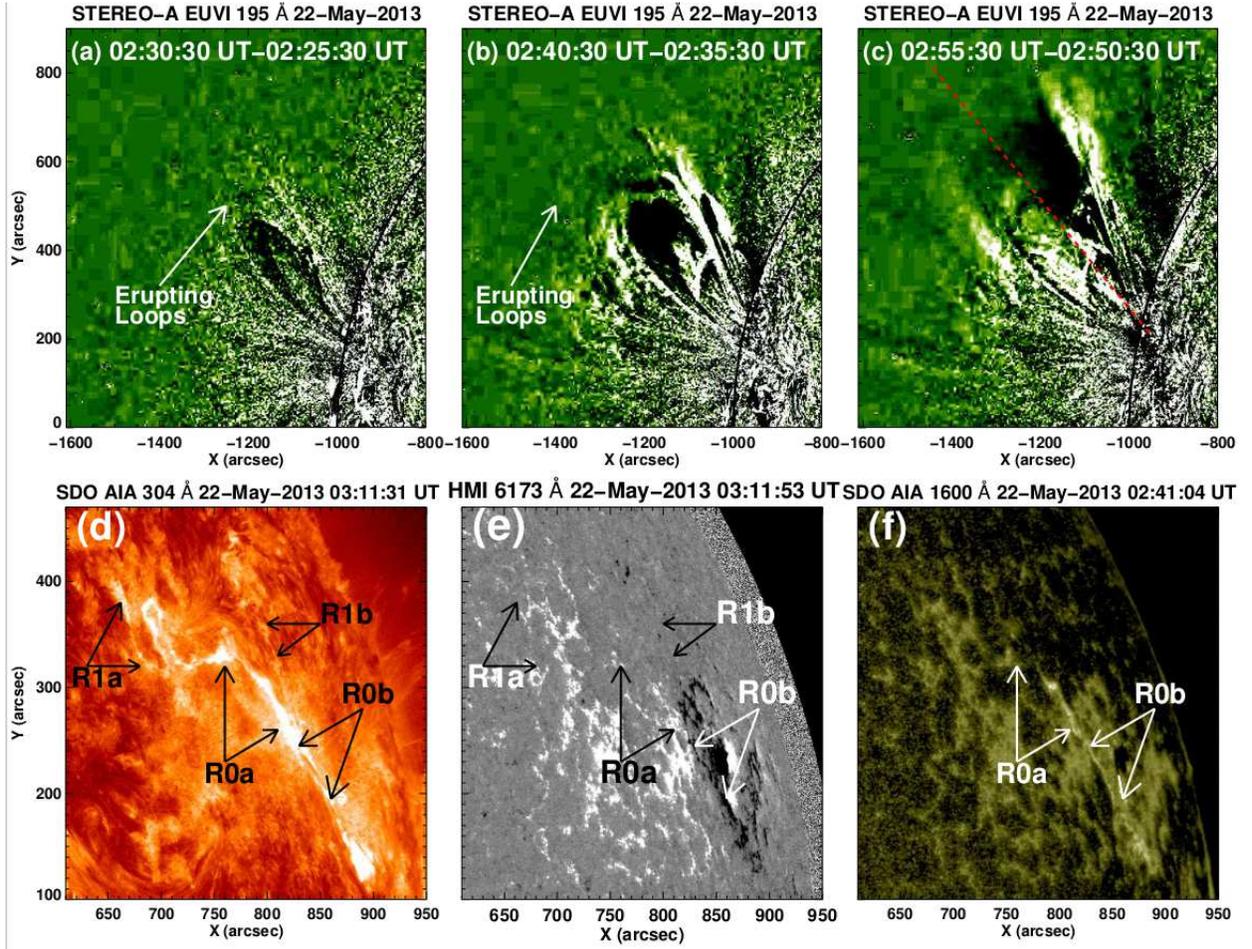}
	}
\vspace*{0cm}
\caption{((a)--(c)) STEREO-A EUVI 195 \AA\ running difference images, showing the blowing--out coronal loops during first CME eruption. The red line in the panel (c) is the path along which the trajectory of the projected height of the erupting structure was measured (see Figure~\ref{fig6}(a)). (d) \textit{SDO}/AIA 304 \AA\ image at 03:11:31 UT showing the flare ribbons of the first CME eruption. (e) \textit{SDO}/HMI line--of--sight magnetogram at 03:11:53 UT on 2013 May 22. Comparison between panels (d) and (f) with (e) show that ribbons R0a and R1a are in positive flux, and ribbons R0b and R1b are in negative flux. An animation of the STEREO-A EUVI 195 \AA\ running difference images showing the eruptions from $\approx$01:15 UT to $\approx$16:55 UT on 2013 May 22 is available in the online journal. Animations of the \textit{SDO}/AIA 304 and 1600 \AA\ images showing the ribbon formation during first eruption from $\approx$02:00 UT to $\approx$03:30 UT on 2013 May 22 are available in the online journal.}
\label{fig5}
\end{figure}

The confined eruption that triggers the first CME eruption starts at $\approx$02:25 UT, with the activation and confined eruption of an upper strand of the low--lying filament. Figure~\ref{fig4} presents \textit{SDO}/AIA 335, 131 and 94 \AA\ images showing this low-lying activity. The activated filament can be seen very well in AIA 335 \AA-channel images at $\approx$02:31 UT (shown by the arrow in Figure~\ref{fig4}(a) and accompanying AIA 335 \AA\ movie). In this confined eruption, this low-lying filament flux rope was arrested within the coronal field of the AR, as can be seen in AIA 131 \AA\ images (shown by the arrow in Figure~\ref{fig4}(b)). The hot flare loops below the arrested flux rope are also observed in the hot AIA 94 \AA\ channel (see Figure~\ref{fig4}(c) and accompanying animation), and are presumably rooted in the inner parts of the two flare ribbons, the parts closely bracketing the PIL (we will refer to these two inner parts of the pair of flare ribbons as R0a and R0b). The ribbons R0a and R0b are pointed to by arrows in Figures~\ref{fig5}(d)--(f), and are seen in the AIA 304 and 1600 \AA\ animations accompanying Figure~\ref{fig5}. This eruption was faint in other AIA channels. This low-lying activity, we believe, triggered the ejective eruption of an overlying flux rope, even though this low-lying filament eruption remained confined in the coronal field of the AR\@. The field that overarches the ejectively erupting flux rope, and that is blown out by it, can be seen in the \textit{SDO}/AIA 94 \AA\ running difference movies accompanying Figure~\ref{fig4}.

The eruption of the overlying coronal loops can also be seen clearly in \textit{STEREO}-A/EUVI 195 \AA\ running difference images (shown by the white arrows in Figures~\ref{fig5}(a)--(c) and accompanying STEREO EUVI 195 \AA\ movies). STEREO-A EUVI observed this eruption from the west, giving us a different viewing angle. We also observed the brightening in the remote regions (marked by R1a and R1b in Figure~\ref{fig5}(d)), which we identify as the flare ribbons made by reconnection below an erupting flux rope. On comparing the ribbons of Figures~\ref{fig5}(d) with the line--of--sight photospheric magnetogram (Figure~\ref{fig5}(e)), we find that R0a and R1a lie in positive flux, and R0b and R1b lie in negative flux.

\begin{figure}[!ht]
\vspace*{0cm}
\centerline{
	\hspace*{0.0\textwidth}
	\includegraphics[width=1\textwidth,clip=]{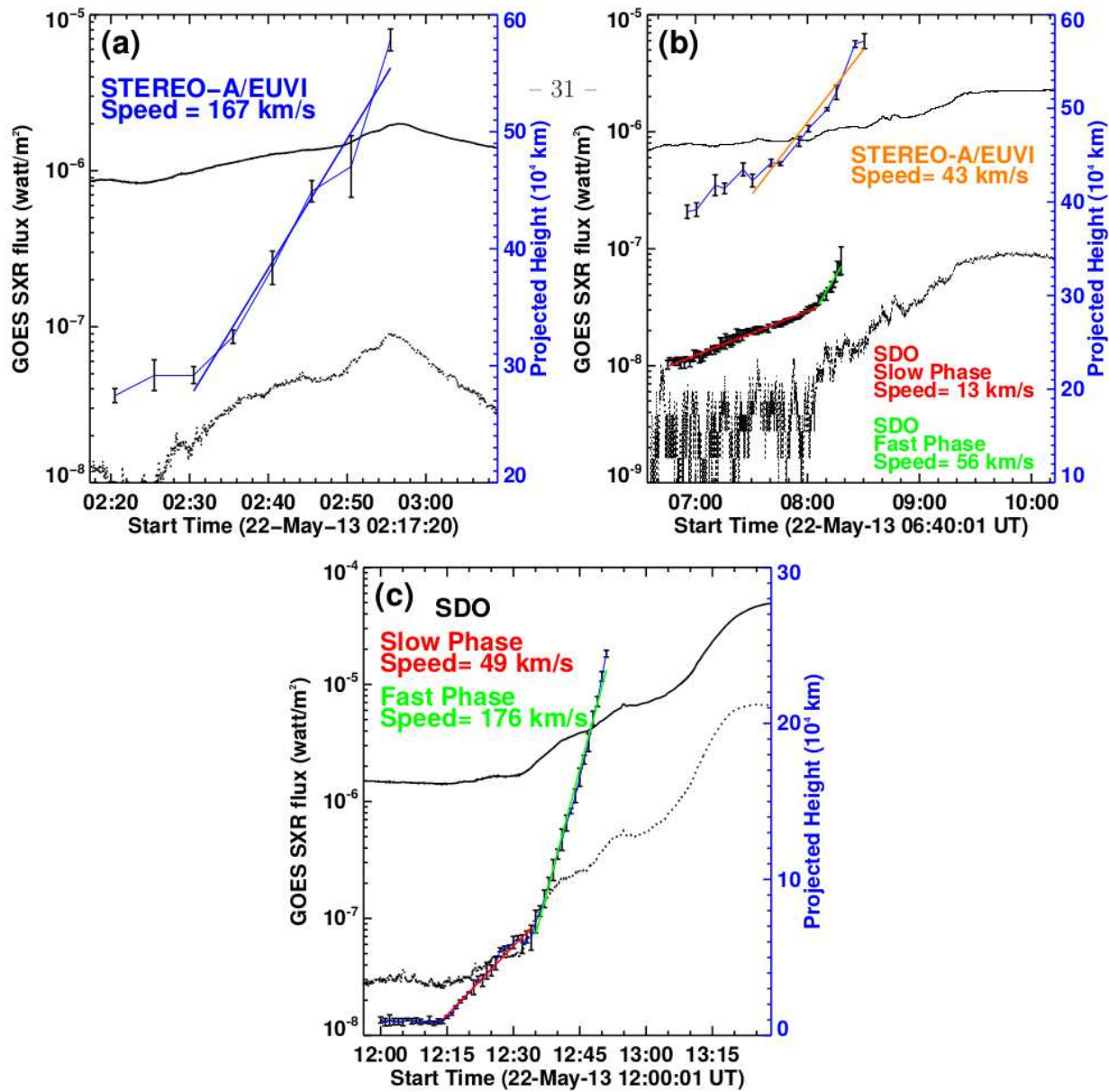}
	}
\vspace*{0cm}
\caption{Projected height--time profiles of the first (a), second (b) and the third (c) CME eruptions using \textit{SDO}/AIA as well as the STEREO-A EUVI images. The GOES X--ray profiles of the 1.0--8.0 \AA\ (solid black) and 0.5-4.0 \AA\ (dotted black) channels are also overplotted for comparison. The speeds are estimated by the linear fits to the height--time data points. Error bars are the standard deviations estimated using the three repeated measurements.} 
\label{fig6}
\end{figure}

The projected height--time plot of this CME eruption using STEREO-A EUVI images is shown in Figure~\ref{fig6}(a). The height--time data points are measured along the dashed red line shown in Figure~\ref{fig5}(c). To obtain the height--time data points, we trace the visible leading edge of the erupting structure. Three repeated measurements of the height data points have been made to get the average value as well as the errors in the height measurements. The average of these three measurements is the value for each point. The error values in the measurements are the standard deviations of the three height measurements. The linear fit to the height--time points gives us the speed, which comes out to be $\approx$167~km~s$^{-1}$ (see Figure~\ref{fig6}(a)).


\subsection{Onset and progression of the second CME eruption}
\label{sec4.3}

\begin{figure}[!ht]
\vspace*{0cm}
\centerline{
	\hspace*{0\textwidth}
	\includegraphics[width=1\textwidth,clip=]{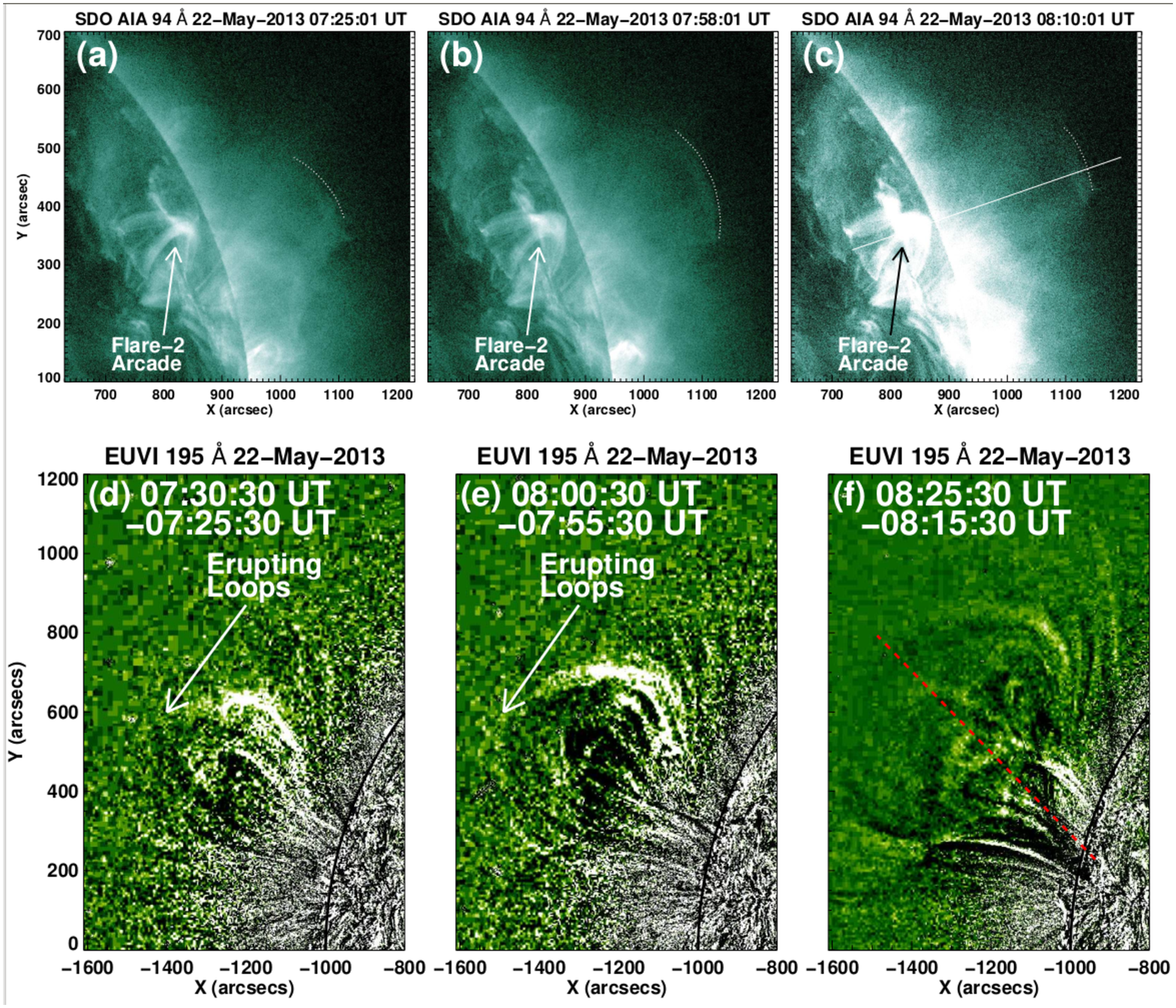}
	}
\vspace*{0cm}
\caption{((a)--(c)) Sequence of selected \textit{SDO}/AIA 94 \AA\ images showing the second CME eruption and the formation of its flare arcade. The white straight line in panel (c) is the path along which the height--time data points were obtained (see Figure~\ref{fig6}(b)). ((d)--(f)) STEREO-A EUVI 195 \AA\ running difference images, showing the blowing--out coronal loops during the second CME eruption. The red line in the panel (f) represents the path along which the projected height points were measured. Animations of the \textit{SDO}/AIA 94 \AA\ intensity (animation 1) and running difference images (animation 2) showing the second and third eruptions from $\approx$05:30 UT to $\approx$13:29 UT on 2013 May 22 are available in the online journal.}
\label{fig7}
\end{figure}

The second CME eruption is observed well in the hot \textit{SDO}/AIA 94 \AA\ channel, as well as in the STEREO-A/EUVI 195 \AA\ channel. A selection of AIA 94 \AA\ images showing the eruption sequence is presented in Figures~\ref{fig7}(a)--(c). Eruption of the visible coronal field lines (shown by the dotted arc in Figures~\ref{fig7}(a)--(c)) starts at around 07:00 UT, and is observed until 08:15 UT in the AIA field--of--view (see the \textit{SDO}/AIA 94 \AA\ intensity and running difference animations accompanying Figure~\ref{fig7}). The eruption was also  observed by STEREO EUVI from the west. The field--loop blowout eruption is clearly seen in EUVI 195 \AA\ running difference images (see Figure~\ref{fig7}(d)--(f) and animation accompanying Figure~\ref{fig5}). The formation of the flare arcade is observed to take place simultaneously with the eruption. The arcade formed just below the erupting structure (see Figures~\ref{fig7}(a)--(c)). The flare arcade started forming at $\approx$06:41 UT\@. This shows that the flux rope in this eruption already started to erupt before the eruption of the outer loops seen in 94 \AA\ started at $\approx$07:00 UT. The flare arcade grew in height with time (see Figures~\ref{fig7}(a)--(c) and accompanying animation).

\begin{figure}[!ht]
\vspace*{0cm}
\centerline{
	\hspace*{0\textwidth}
	\includegraphics[width=1\textwidth,clip=]{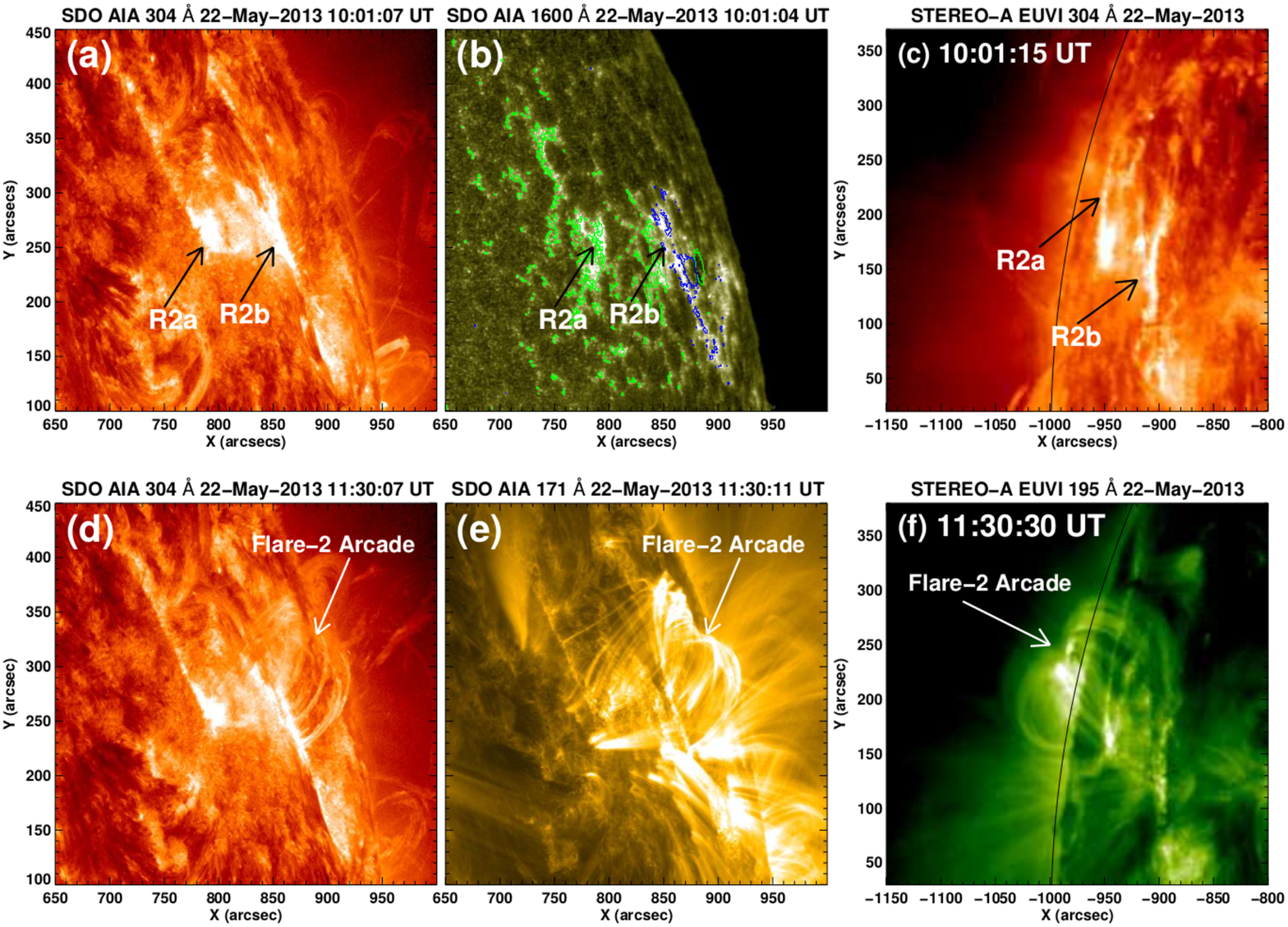}
	}
\vspace*{0cm}
\caption{((a)--(b))\textit{SDO}/AIA 304 \AA\ and 1600 \AA\ images showing the two ribbons of the second CME flare at $\approx$10:01 UT\@. The green and blue contours in panel (b) map the positive and negative flux of the \textit{SDO}/HMI magnetogram. The contours levels are $\pm 100$ Gauss. (c) STEREO-A EUVI 304 \AA\ image at $\approx$10:01 UT showing the two ribbons of the second CME flare. ((d)--(e))\textit{SDO}/AIA 304 \AA\ and 171 \AA\ images, showing the flare arcade of the second CME flare during the late decay phase $\approx$11:30 UT. (f) STEREO-A EUVI 195 \AA\ images, also showing the flare arcade at $\approx$11:30 UT\@.} 
\label{fig8}
\end{figure}

The two feet of the flare arcade are in the two flare ribbons (Figures~\ref{fig8}(a)--(b)). Figures~\ref{fig8}(a) and~\ref{fig8}(b) display \textit{SDO}/AIA 304 \AA\ and 1600 \AA\ images at around $\approx$10:01 UT, showing the flare ribbons near the western limb.  HMI photospheric magnetogram contours (green and blue for positive and negative polarities, respectively) are overplotted onto the \textit{SDO}/AIA 1600 \AA\ image (Figure~\ref{fig8}(b)). The eastward and westward ribbons (marked by R2a and R2b, respectively, in Figures~\ref{fig8}(a)--(b)) are in the positive- and negative-polarity regions, respectively. The flare ribbons can also be seen with the STEREO-A EUVI 304 \AA\ channel, 
viewed from the west (Figure~\ref{fig8}(c)). The flare arcade during the late decay phase at $\approx$11:30 UT can be seen in 304 \AA\ (Figure~\ref{fig8}(d)) and 171 \AA\ (Figure~\ref{fig8}(e)) of AIA as well as in the STEREO-A EUVI 195 \AA\ (Figure~\ref{fig8}(f)).

The height--time profile of the second CME eruption is shown in Figure~\ref{fig6}(b). The trajectory along which the height--time data points are obtained is shown in Figure~\ref{fig7}(c) with the solid white line. From the AIA 94 \AA\ height--time profile, we observe two steps of evolution: an initial slow eruption with an average speed of around $\approx$13~km~s$^{-1}$ from $\approx$06:45 UT to 
$\approx$08:00 UT (see red line in Figure~\ref{fig6}(b)), and a second stage during which it accelerates to $\approx$56~km~s$^{-1}$ from $\approx$08:00 UT to $\approx$08:15 UT (see green fitted line in Figure~\ref{fig6}(b)). The height--time profile of STEREO EUVI 195 \AA\ show the same two
steps, and gives  $\approx$43~km~s$^{-1}$  for the speed in the second phase (see the orange line in Figure~\ref{fig6}(b)). Such two-stage rises in filament trajectories are not uncommon \citep[][]{Sterling5,McCauley15,Joshi16a,Mitra19}. 



\subsection{Onset and progression of the third CME eruption}
\label{sec4.4}

\begin{figure}[!ht]
\vspace*{0cm}
\centerline{
	\hspace*{0\textwidth}
	\includegraphics[width=1\textwidth,clip=]{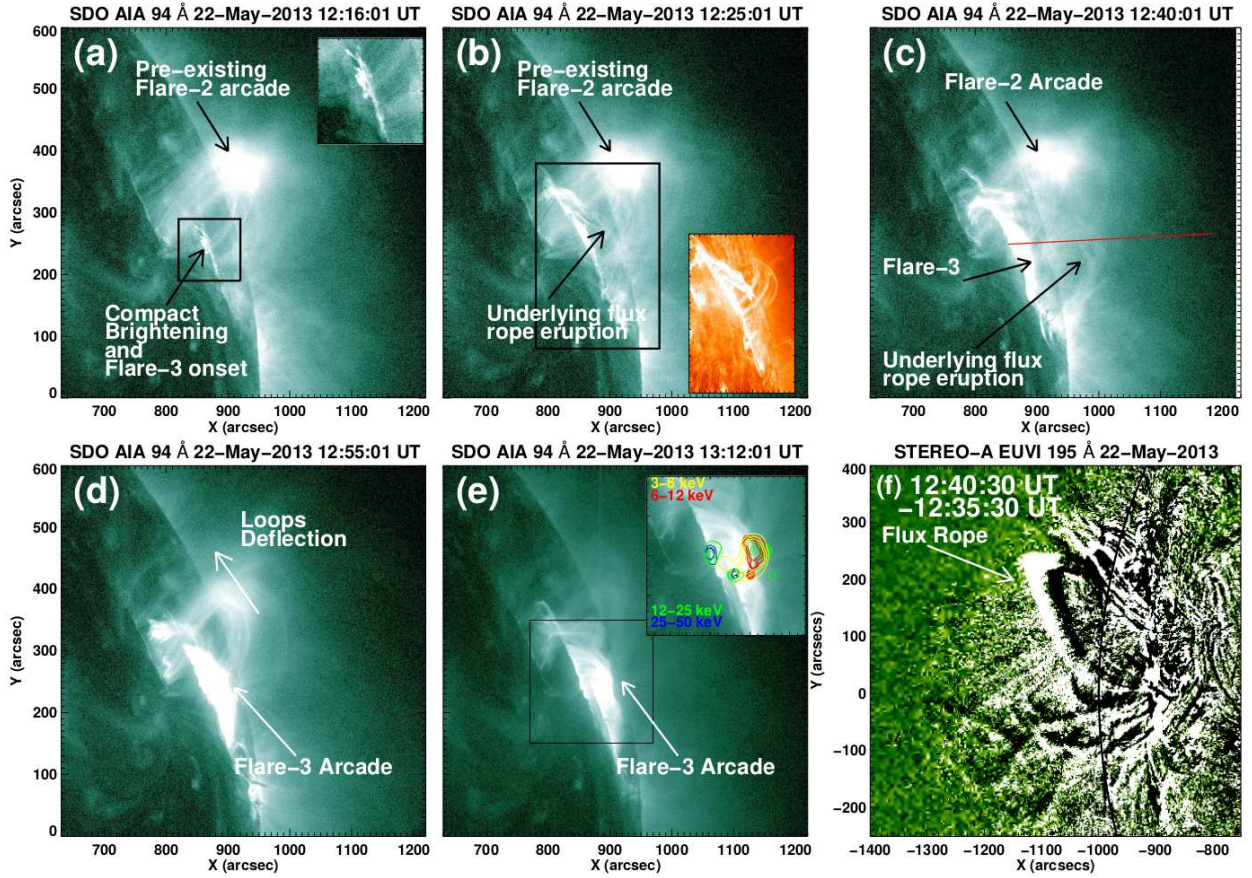}
	}
\vspace*{0cm}
\caption{((a)--(e)) Sequence of \textit{SDO}/AIA 94 \AA\ images showing the flux rope eruption early in the third CME eruption and the formation of the resulting flare arcade. The inset in panel (a) is a image of the FOV outlined by the black box in panel (a). The inset in panel (b) is a simultaneous 304 \AA\ image of the FOV outlined by the black box in the 94 \AA\ image. The red straight line in panel (c) is the trajectory along which the height-time data points were obtained (see Figure~\ref{fig6}(c)). The RHESSI X-ray sources are shown in the inset in panel (e). (f) STEREO-A EUVI 195 \AA\ image at $\approx$12:40 UT, showing the erupting flux rope during the third eruption.} 
\label{fig9}
\end{figure}

The third CME eruption was well observed in almost all EUV channels of AIA\@. SDO/AIA 94 \AA\ images showing the eruption of the flux rope are presented in Figure~\ref{fig9}(a)--(c). This episode starts with a compact brightening near the polarity inversion line at around $\approx$12:13 UT (shown by the arrow in Figure~\ref{fig9}(a)). This brightening is probably due to runaway tether--cutting reconnection between the legs of the low-lying sheared arcade. The flux rope eruption starts at $\approx$12:13 UT (Figure~\ref{fig6}(c)) and was in its fast-rise phase by $\approx$12:40 UT (Figure~\ref{fig6}(c)). The compact brightening and the onset of the eruption of the flux rope take place under the second-flare arcade (see Figure~\ref{fig9}(a)--(c) and the animation accompanying Figure~\ref{fig7}). The reconnection takes place below the erupting flux rope between the legs of the enveloping arcade. As a result, the flare arcade is formed underneath the erupting flux rope and can be seen in Figures~\ref{fig9}(d)--(e). We also observed RHESSI X--ray sources at the top of the flare arcade as well as at the feet (see different color contours in the inset of Figure~\ref{fig9}(e)). The erupting flux rope can also be seen in STEREO-A EUVI 195 \AA\ running difference image at $\approx$12:40 UT (Figure~\ref{fig9}(f) and the STEREO-A EUVI 195 \AA\ running difference animation accompanying Figure~\ref{fig5}). The erupting flux rope blows out the overlying arcade that formed during the second CME eruption (Figure~\ref{fig9}(d)).

\begin{figure}[!ht]
\vspace*{0cm}
\centerline{
	\hspace*{0\textwidth}
	\includegraphics[width=1\textwidth,clip=]{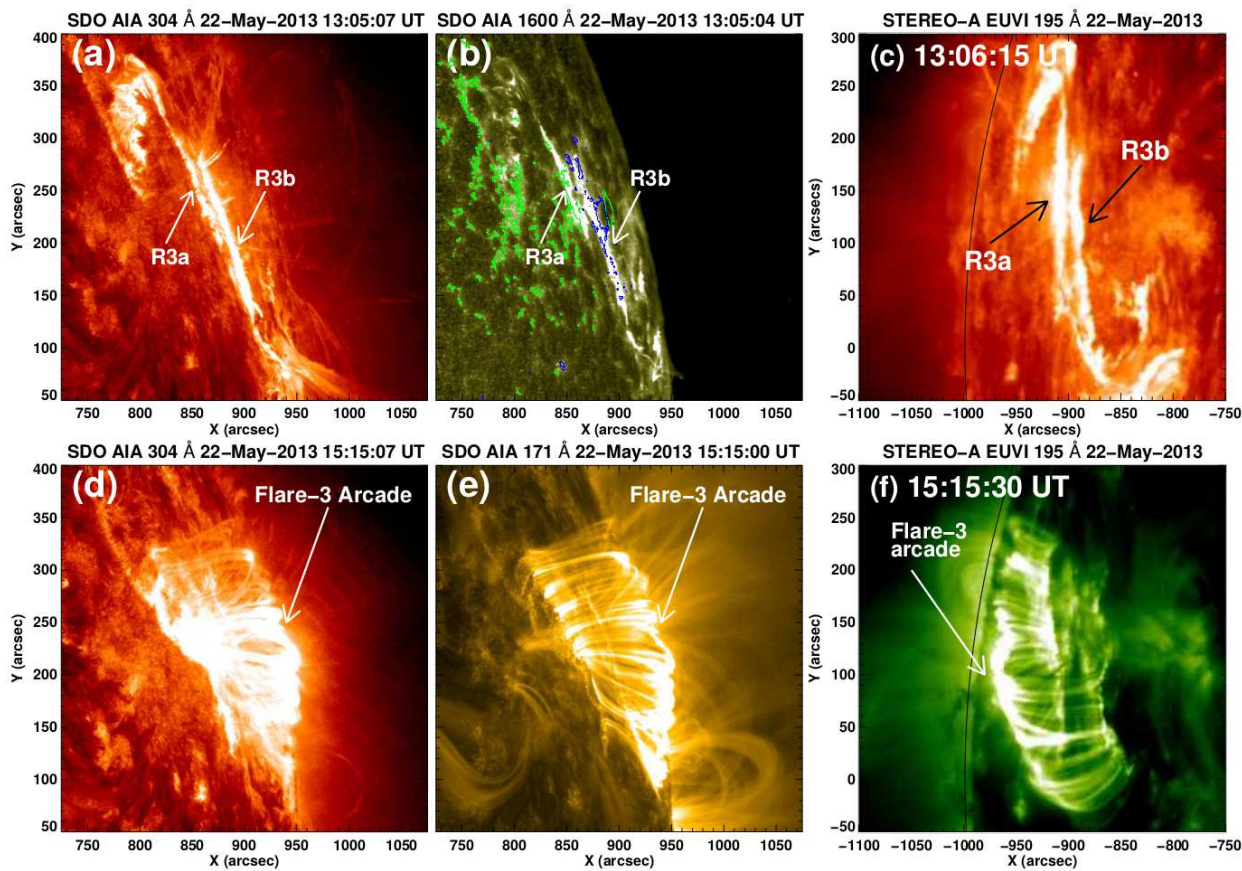}
	}
\vspace*{0cm}
\caption{((a)--(b))\textit{SDO}/AIA 304 \AA\ and 1600 \AA\ images, showing the two ribbons of the third CME flare at $\approx$13:05 UT viewed from the east. (c) STEREO-A EUVI 304 \AA\ at $\approx$13:06 UT, also showing the third-CME flare ribbons viewed from the west. ((d)--(e))\textit{SDO}/AIA 304 \AA\ and 171 \AA\ images showing the late--phase flare arcade of the third CME flare at around $\approx$15:15 UT. (f) STEREO-A EUVI 195 \AA\ image at $\approx$15:15 UT, also showing the late--phase flare arcade of the third CME flare.} 
\label{fig10}
\end{figure}

This eruption produced two parallel flare ribbons seen in AIA (see Figure~\ref{fig10}(a)--(b)) and STEREO-A EUVI (see Figure~\ref{fig10}(c)) images. The eastern and western ribbons of the flare are marked as R3a and R3b, respectively, in Figure~\ref{fig10}(a)--(c). HMI photospheric line--of--sight contours are overplotted on the \textit{SDO}/AIA 1600 \AA\ image with green (positive) and blue (negative) colors. The eastern and western ribbons lie on the positive and negative polarity regions, respectively. The flare arcade connecting the two ribbons can be seen at the later time at $\approx$15:15 UT in AIA 304 \AA\ (Figure~\ref{fig10}(d)), 171 \AA\ (Figure~\ref{fig10}(e)) and STEREO EUVI 195 \AA\ (see Figure~\ref{fig10}(f)) images.

\subsection{Accompanying CMEs}
\label{sec4.5}

Figure~\ref{fig11} presents the evolution of three successive CME eruptions, as observed in both STEREO-A COR1 and LASCO C2 and C3 fields of view. A weak CME was made by the first CME eruption (Figures~\ref{fig11}(a)--(c)). This first CME appears in the STEREO COR1 field--of--view at $\approx$02:55 UT (Figure~\ref{fig11}(a)) and in the LASCO C2 field of view at $\approx$03:15 UT (Figure~\ref{fig11}(c)). This first CME only shows a leading-edge structure without a visible CME core. The second and third CMEs on the other hand, show full three-part structure (see Figures~\ref{fig11}(d)--(f) and Figures~\ref{fig11}(g)--(i), respectively). This is evidence that the second and third CMEs were made by the eruption of flux ropes, which appear as the CME cores in the second and third CMEs (see Figures~\ref{fig11}(e) and Figures~\ref{fig11}(i), respectively) The speeds of the first, second and third CMEs are $\rm 244~km~s^{-1}, 687~km~s^{-1} and~1466~km~s^{-1}$, respectively\footnote{\url{https://cdaw.gsfc.nasa.gov/CME_list/UNIVERSAL/2013_05/univ2013_05.html}}.


\section{Results and Discussion}
\label{sec5}

In this work, we analyzed the sequence of four consecutive eruptions that occurred on 2013 May 22 from a single AR, NOAA AR 11745. The active region was overall bipolar with a straight PIL. We interpreted the progression of the eruption sequence using a multi-wavelength and multi-view data set from \textit{SDO} and \textit{STEREO-A}. The main findings of this work are as follows.

\begin{enumerate}
\item Our observations are consistent with the scenario that initially there was a stack of three flux ropes above the PIL of the AR\@. We call this configuration a ``triple-decker" flux rope configuration.

\item The eruption of first (highest) flux rope is triggered by the activation and confined eruption of the middle flux rope, which is 
an upper strand of a low-lying filament.  This confined eruption disturbs the overlying field lines and triggers the CME-producing eruption of the topmost flux rope.  The confined eruption made flare ribbons close to the PIL, and the CME eruption produced ribbons far-offset from the PIL\@.

\item The second CME eruption is the blowout eruption of the previously--arrested middle flux rope. The removal of overlying flux by the first flux rope's eruption 
triggers this second CME-producing flux rope eruption. A large scale flare arcade and ribbons are also observed to be made by this second eruption.

\item In the third CME eruption the lowest flux rope (carrying the remaining strands of the filament) blows out, followed by the formation of a new flare arcade and flare ribbons. The removal of overlying flux by the second CME eruption triggered this third CME eruption.

\item The first CME shows only faint structure, while the second and third CMEs show full 
three-part structure. We presume that there also was an erupting flux rope inside the first CME, although the flux-rope structure of the
second and third CMEs is more obvious.

\item The sequential removal of overlying flux \citep[i.e., lid removal;][]{Sterling14} during the first and second CME eruptions respectively trigger the second and third CME eruptions. We suppose that the confined first eruption that triggered the first CME was triggered by slow flux--cancellation tether-cutting reconnection at the PIL\@ \citep[][]{Moore92}.

\begin{figure}[!ht]
\vspace*{0cm}
\centerline{
	\hspace*{0\textwidth}
	\includegraphics[width=1\textwidth,clip=]{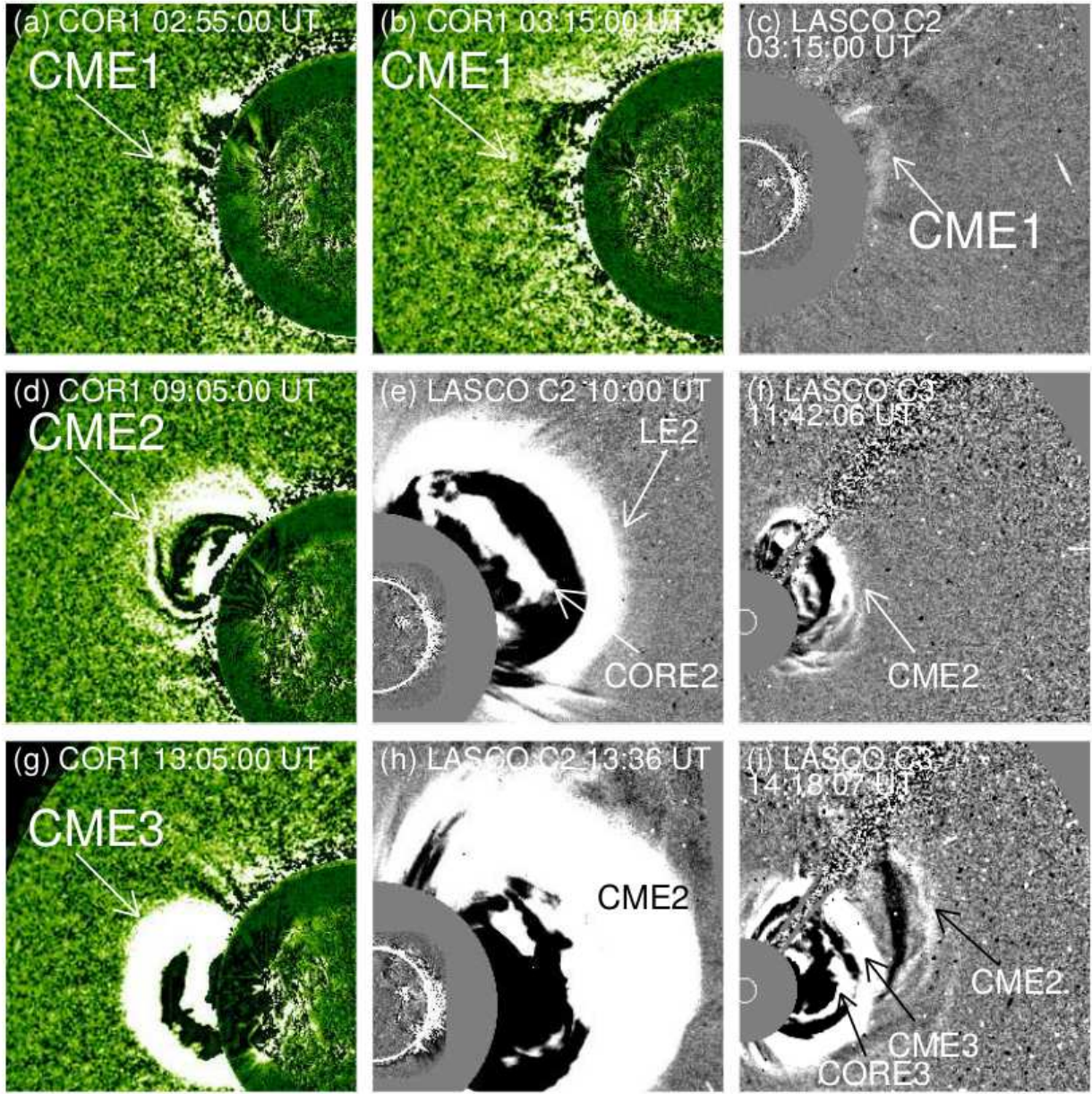}
	}
\vspace*{0cm}
\caption{STEREO-A COR1 ((a)--(b)) and LASCO C2 (c) images, showing the evolution of the CME made by the first eruption. STEREO-A COR1 (d), LASCO C2 (e) and LASCO C3 (f) images, showing the evolution of the CME made by the second eruption. STEREO-A COR1 (g), LASCO C2 (h), and LASCO C3 (i) images, showing the evolution of the CME made by the third eruption. The LASCO difference images are taken from the LASCO-CME catalog $\rm https~:~//cdaw.gsfc.nasa.gov/CME\_list/UNIVERSAL/2013\_05/univ2013\_05.html$.}
\label{fig11}
\end{figure}

\end{enumerate}

\begin{figure}
\centerline{
	\hspace*{0.0\textwidth}
	\includegraphics[width=1\textwidth,clip=]{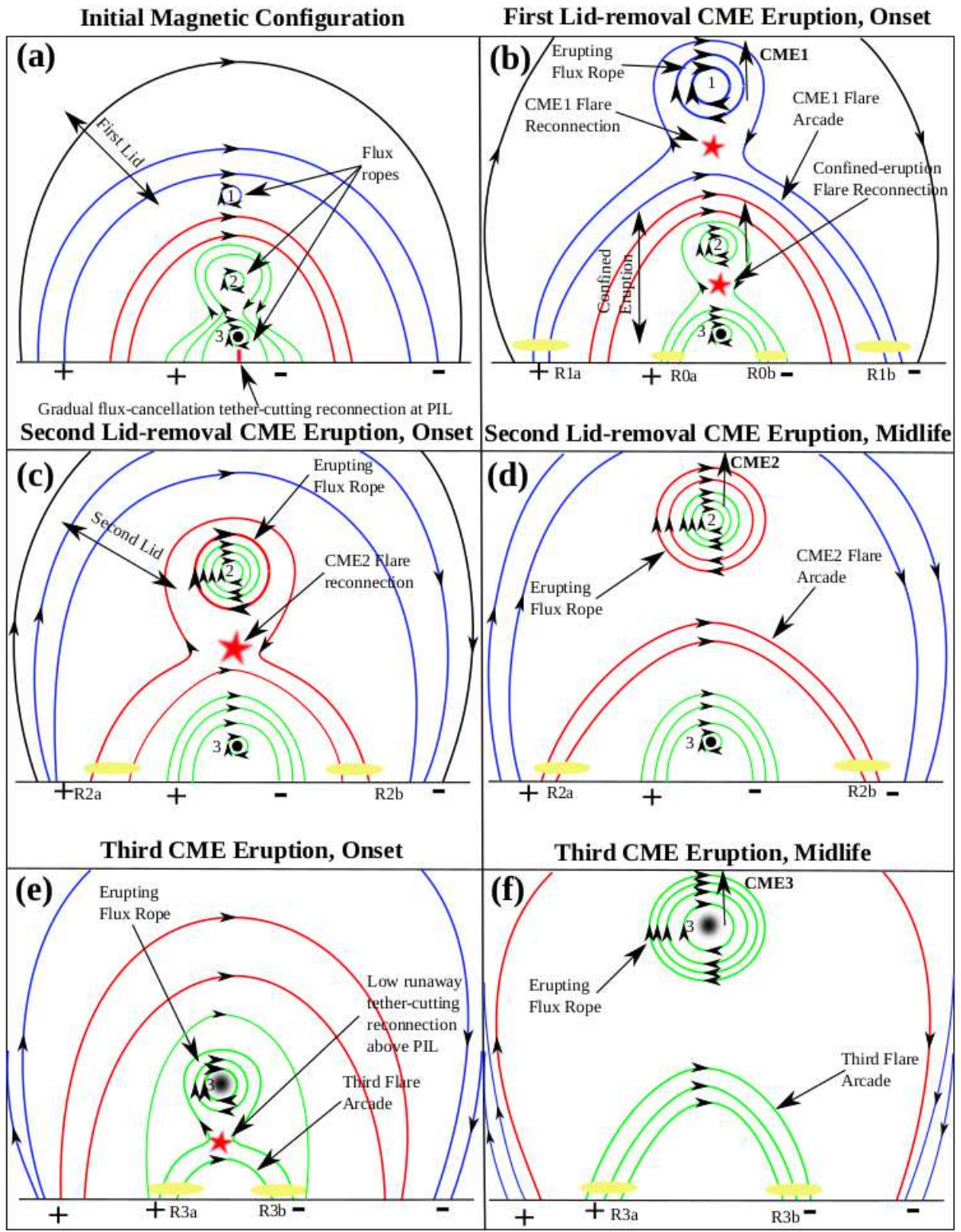}
	}
\vspace*{0cm}
\caption{Schematic of the sequence of the eruptions based on the multi-wavelength observations.}
\label{fig12}
\end{figure}

A schematic representation of this sequence of eruptions is shown in Figure~\ref{fig12}. 
In Figure~\ref{fig12}(a), the coronal loops from low to high heights are shown by green, red, blue and black colors respectively, along with flux ropes in between. The first CME eruption is triggered by the activation and confined eruption of upper strands of the filament in the middle flux rope, i.e.\ flux rope 2 (Figure~\ref{fig12}(b) and ~\ref{fig4}(a)).
This confined eruption is plausibly triggered by slow flux-cancellation tether--cutting  reconnection at the PIL \citep[][]{Moore92} (see HMI magnetic field animations accompanying Figure~\ref{fig2}). This confined eruption of the flux rope 2 filament disturbed the equilibrium of the topmost flux rope 1 and triggered it to erupt (cf.\ Figure~\ref{fig12}(b) and~\ref{fig5}(a)--(c)). 
Flare reconnection from the confined eruption occurred underneath the erupting flux rope (shown by the lower red star in Figure~\ref{fig12}(b)) causing heated innermost flare ribbons denoted as R0a and R0b (Figures~\ref{fig5}(d)--(f)). The continued eruption of the topmost flux rope 1 is followed by the reconnection underneath it, heating the farther-separated pair of flare ribbons (R1a and R1b  in Figures~\ref{fig12}(b) and~Figures~\ref{fig5}(d)--(e)) and unleashing the first CME (cf.\ Figure~\ref{fig12}(b) and~\ref{fig11}(a)--(c)). At this time flux rope 2 does not erupt, but remains stable at some higher height in the solar corona for a few hours. It then starts to erupt upward slowly and drives CME 2 flare reconnection underneath it between the red field lines (Figure~\ref{fig12}(c) and Figure~\ref{fig7}). This makes a new pair of flare ribbons (see yellow ovals in Figures~\ref{fig12}(c)--(d) and Figures~\ref{fig8}(a)--(c)) and flare arcade (cf.\ Figure~\ref{fig12}(d) and Figures~\ref{fig8} (d)--(f)). The erupting flux rope 2 drives the second CME (CME2 in Figure~\ref{fig12}(d) and Figures~\ref{fig11}(d)--(f)). The third CME eruption starts with low--lying runaway tether--cutting reconnection (cf.\ red star in Figure~\ref{fig12}(e) and compact brightening in Figure~\ref{fig9}(a)). The filament and enveloping flux rope then erupt and drive further flare reconnection underneath. The ribbons (shown by the yellow colored ovals in Figures~\ref{fig12}(e)--(f) and Figures~\ref{fig10}(a)--(c)) and the flare arcade (Figures~\ref{fig9}(e)--(f) and Figures~\ref{fig10}(d)--(f)) form as a result of this reconnection. The erupting flux rope blows out the third CME (CME3 in Figure~\ref{fig12}(f) and~Figure~\ref{fig11}(g)--(i)).

We refer to the stack of three flux ropes above the PIL as a ``triple-decker configuration."  To our knowledge, our work here
is the first proposing the presence of this triple-decker magnetic configuration over the PIL of a bipolar AR from observation and analysis. However, earlier there were some studies where the evolution of a double-decker filament system has been observed \citep[e.g.,][]{Liu12b,Kliem14,Zhu14}. Those studies suggested two types of double-decker filament configurations: in the first configuration, there is a flux rope above the sheared arcades with dips, while in the second configuration two flux ropes lie stacked over the AR's PIL, i.e.\ one over the other \citep[see Figure~\ref{fig12} of][]{Liu12b}. In our case, i.e., the triple-decker configuration, all three of the flux ropes lie stacked above the AR's PIL\@. The formation of flare ribbons and flare arcades across the same PIL after each eruption confirms the presence of this magnetic configuration.

It is important to learn how these typical double-decker and triple-decker flux rope magnetic configurations form. Two mechanisms for the formation of double-decker filaments has been discussed by \citet{Liu12b}. In the first possibility the upper branch first 
forms over the PIL, and the lower branch emerges later to form the double-decker filament system. A possible such emergence of a flux rope under a pre-existing filament has already been observed \citep[][]{Okamoto08}. In the second possibility, initially both branches of the double-decker filament system reside close together, and later are separated by a partial eruption \citep[][]{Gilbert01}. In this scenario, the reconnection within the filament flux rope splits the filament flux rope into two branches. In the present study, our observations suggest the presence of a more complex triple-decker flux rope configuration.  We speculate that the flux ropes may have formed successively via tether cutting at the PIL\@.  We suppose each 
flux rope reached some stable coronal height during a confined eruption and remained there until further eruption. In this supposed 
process, the stack of three flux ropes over one another was formed.

Various mechanisms have been proposed to explain the onset of solar eruptions, e.g. runaway tether cutting \citep{Moore01}, magnetic breakout \citep{Antiochos99} and flux emergence \citep{Chen00}. All these mechanisms are restricted to the erupting field only. In the runaway tether--cutting mechanism, the reconnection begins underneath the flux rope, internal to the erupting field. In the breakout
scenario the reconnection begins at the top, external to the erupting field. In the flux--emergence reconnection mechanism, flux cancellation between existing and emerging nearby field triggers the eruption. The eruption of a double-decker filament due to the interaction and merging between the two branches has been proposed in some studies \citep[][]{Zhu14,Liu12b,Kliem14}. They suggested that the transfer of mass, magnetic flux and current from lower branch to upper branch during the interaction is the key mechanism for the upper branch to lose equilibrium. Recently, a new mechanism, referred to as the lid-removal mechanism \citep{Sterling14}, has been investigated, where it is proposed that the removal of some of the overlying field (blow out of the top of the envelope field) destabilizes the tension--pressure balance in the underlying field. In our case, we observed successive removal of overlying field during the first and second CME eruptions that apparently trigger the second and third CME eruptions, respectively. Thus, our observations support the lid--removal scenario.

The stability of double-decker filaments using MHD simulation is discussed by \citet[][]{Kliem14}. In their configuration, they considered two vertically arranged force--free flux ropes in a bipolar enveloping external field. In their simulation, they found that the external field's toroidal (shear) component is a key parameter controlling the stability of this configuration if the two flux ropes lie close to each other. The double-decker configuration remains stable if the strength of the external toroidal (shear) field is high. In the case where the toroidal component decreases sufficiently, both flux ropes erupt with the initial eruption of the lower flux rope. In the case when the shear field is above a threshold value, the configuration undergoes a partial eruption of the higher flux rope. We also expect that the external magnetic field strength is a key factor for stability or eruption in our triple-decker flux rope configuration. Numerical simulation of such a triple-decker configuration is required to understand more about the stability/eruption conditions in such configurations.

So far, it is unclear how common these types of sequential eruptions from a single bipolar active region may be. 
The operation of the lid--removal mechanism needs to be modeled in order to understand fully the eruption scenario. Further observational and numerical simulation studies are required to improve our insight into triggering of eruptions by lid removal.


\acknowledgments
We thank SDO/AIA, SDO/HMI, GOES, STEREO, and RHESSI teams for providing their data for the present study. A.C.S. and R.L.M. received funding from the Heliophysics Division of NASA’s Science Mission Directorate through the Heliophysics Guest Investigators (HGI) Program. We thank the referee for his/her valuable and constructive comments and suggestions.



\end {document}